\documentclass[twocolumn,amsmath,amssymb,pre]{revtex4}
\usepackage{graphicx,color,longtable}

\begin{document}

\makeatletter

\title{In Search of Colloidal Hard Spheres}

\author{C. Patrick Royall}
\affiliation{School of Chemistry, University of Bristol, Bristol, BS8
1TS, U.K and International Research Center for Materials Nanoarchitectonics (MANA), National Institute for Materials Science (NIMS), Tsukuba, Ibaraki 305-0044, Japan.}

\author{Wilson C. K.  Poon}
\affiliation{Scottish Universities Physics Alliance (SUPA) and The School
of Physics, University
of Edinburgh, Kings Buildings, Mayfield Road, Edinburgh EH9 3JZ,
U. K.}

\author{Eric R. Weeks}
\affiliation{Department of Physics, Emory University, Atlanta, GA
30322 U.S.A.}

\begin{abstract}
We recently reviewed the experimental determination of the volume
fraction, $\phi$, of hard-sphere colloids, and concluded that the
absolute value of $\phi$ was unlikely to be known to better than $\pm
3$-6\%. Here, in a second part to that review, we survey effects
due to softness in the interparticle potential, which necessitates
the use of an {\em effective} volume fraction. We review current
experimental systems, and conclude that the one that most closely approximates
hard spheres remains polymethylmethacrylate spheres
sterically stabilised by polyhydroxystearic acid `hairs'. For these
particles their effective hard sphere diameter is around 1-10\%
larger than the core diameter, depending on the particle size. We
argue that for larger colloids suitable for confocal microscopy,
the effect of electrostatic charge cannot be neglected, so that
mapping to hard spheres must be treated with caution.
\end{abstract}
\vspace{0.5cm}
  
\maketitle





\section{Introduction}

A collection of hard spheres is one of the simplest examples of an interacting system.
Hard-sphere packings have been important since the dawn of civilisation 
\cite{hales1997i,hales1997ii}, while using hard spheres to model the liquid state
dates back at least to Kirkwood \cite{kirkwood1942} in the 1940s. 
The simplicity of hard spheres lends itself to analytical theory \cite{werthiem1963,thiel1963,lebowitz1964,mansoori1971}
and computer simulation.\cite{rosenbluth1954,alder1957,wood1957}  
While the absence of inter-particle attraction precludes a liquid phase,  concentrated hard
sphere fluids capture many properties of the liquid state.
\cite{kirkwood1942,widom1967,weeks1971}  Due to its analytic tractability, and
that it is described by just one parameter, volume fraction
$\phi$, the importance of the hard sphere system as a basic model in understanding
condensed matter cannot be overstated.

Early hard sphere experiments include Bernal's use of ball bearings to
model liquid structure.\cite{bernal1959}   However, ball bearings
have negligible thermal motion. Suspensions of mesoscopic
colloids exhibit Brownian motion and are thermodynamically
equivalent to atoms and small molecules \cite{pusey}. In 1986,
Pusey and van Megen \cite{pusey1986} showed that a suspension
of sterically-stabilised polymethylmethacrylate (PMMA) particles
showed hard-sphere-like equilibrium phase behaviour. Their work
led to many experimental studies of the statistical physics
of hard spheres using colloids as models.  Since Pusey and van
Megen's work, the equation of state of hard-sphere colloids has
been determined \cite{piazza1993}, crystal nucleation has been
observed \cite{aastuen1986,schatzel1992}, and the glass transition
has been studied \cite{vanMegen1998}.

The body of experimental research just reviewed relied on light scattering as the structural and dynamical probe. The advent of single particle tracking in real space with confocal
microscopy \cite{vanBlaaderen1992,vanblaaderen1995} opened a new dimension in 
experiments on hard-sphere-like systems, yielding an unprecedented level of detailed information \cite{prasad2007}.
Confocal microscopy of hard-sphere-like suspensions is thus ideal
for studying generic processes where local events are important, such
as crystal nucleation \cite{gasser2001}, melting \cite{alsayed2005}
and dynamical heterogeneity \cite{kegel2001,weeks2001}. 

In principle, the thermodynamics of a system of hard spheres is controlled solely by the state variable $\phi$. We have recently reviewed the experimental determination of
volume fraction\cite{poon2011} and concluded that, although
relative values of $\phi$ may be known with high precision, absolute
values can only be determined to within 3-6\% accuracy. This matters, especially when dealing  with dynamical properties (e.g., phase transition kinetics and the glass transition), since these can be very strong functions of $\phi$. 

But the accurate determination of $\phi$ is only part of the experimental challenge. The other part of the challenge was hinted at by the title of Pusey and van Megen's 1986 paper, ``Phase behaviour of concentrated
suspensions of \emph{nearly} hard colloidal spheres'' \cite{pusey1986}, where we have added the italics to emphasise the point in question, namely, that true hard spheres do not exist in reality. In Pusey and van Megen's case, the lack of hardness is almost certainly due to the small but finite compressibility of the PHSA stabilising `hairs'. The same is generically true of other sterically-stabilised particle systems. 

In very-nearly-hard, sterically-stabilised suspensions, a new
sample has to be made for every state point $\phi$. Apart from
being cumbersome, this also restricts the accuracy with which the
sole thermodynamic control parameter can be `tuned'. Thus, there
has been a drive to use particles such as `microgels', whose
diameter is temperature dependent. Since temperature, $T$, can be
tuned far more accurately than $\phi$, this allows the scanning
of a single sample very finely through $\phi$ space by varying
$T$. But the price paid for the `tunability' of particle
diameter is that some softness is built in by design.

\begin{figure}
\centering
\includegraphics[width=8cm,keepaspectratio]{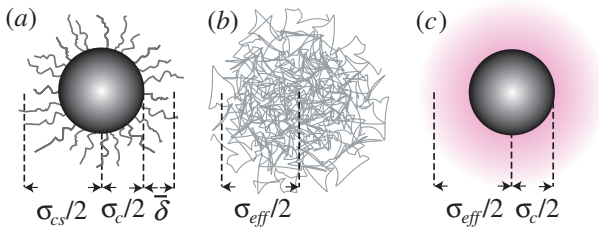} 
\caption{Schematic representation of various models for hard-sphere colloids.  (a) Sterically-stabilized
particle, with surface `hairs' (not to scale), where the average
hair thickness $\bar{\delta}$ and the core-shell diameter $\sigma_{cs}=\sigma_{c}+2\bar{\delta}$ are needed for a full characterisation. (b) Microgel particle, which is a heavily cross-linked polymer.  (c) Charged colloid, where the electrical double layer (shaded) gives rise to an effective diameter $\sigma_{\rm eff}$.}
\label{figRealHard} 
\end{figure}

To understand a final reason why real model hard-sphere colloids
may be somewhat soft, consider why in practice, not only $\phi$
but the particle diameter $\sigma$ matters. Experimentally,
colloids as synthesised are usually not matched in density to the
dispersing medium. Significant sedimentation (or, less usually,
creaming) during the timescale of an experiment therefore
presents a problem. In light scattering studies, this is
circumvented to a large extent by the use of small particles,
say, $\sigma \lesssim 400$nm, so that for PMMA particles
dispersed in cis-decalin, a single particle sediments
$\lesssim1$mm per day. The middle part of a bulk sample would
therefore be little affected by sedimentation over a day. But if
one wants to eliminate the effect all together (e.g. for
long-time measurements), or when larger particles (say, $\sigma
\gtrsim 1000$nm) are needed for imaging purposes, then solvent mixtures for density matching are needed, which often introduces significant charging. The practical need for larger particles therefore inevitably brings softness.  

In this second part of our review, we survey critically these three sources of softness in experimental systems of colloids that have been used to model hard spheres, Fig. \ref{figRealHard}. In the spirit of the first part of our review \cite{poon2011}, we seek to provide a means by which
the hardness may be assessed, and consider the consequences this may
have on the behaviour of the system. Likewise, we argue for more clarity
on the part of experimentalists, concerning the softness of the systems they use. We also suggest a number of criteria by which experimentalists (and theorists using experimental data) may judge whether a certain  system of colloids may be considered `hard enough' for answering particular physics questions. 

Below, we first discuss mapping experimental systems onto hard
spheres via measuring the inter-particle potential in section
\ref{sectionMapping}. We then treat softness of non-electrostatic
origins in section \ref{sectionIntrinsic} and of electrostatic origins in
section \ref{sectionElectrostatics}; in general, both effects
are present simultaneously.  Finally we give a worked example in
section \ref{sectionCaseStudy}, illustrating many of the points
raised throughout this review, before concluding in section
\ref{sectionConclusions}.

\section{Mapping to effective hard spheres via the interaction potential}
\label{sectionMapping}

Given that real colloids inevitably display a degree of softness,
it is important to be able to map their behaviour to that of hard
spheres for the purpose of comparison with theories and simulations
of perfect hard spheres.  By mapping, we mean finding an effective
hard-sphere diameter $\sigma_{\rm eff}$ so that one may map from
the experimentally-controllable particle number density $\rho$
to an effective hard-sphere volume fraction $\phi_{\rm eff}$
using $\phi_{\rm eff}=\pi\rho\sigma_{\rm eff}^{3}/6$.  There are
two conceptually distinct ways of determining $\sigma_{\rm
eff}$. First, one could map via some known hard-sphere property,
such as the volume fraction at freezing ($\phi_f^{\rm HS} = 0.494$), or the
viscosity of the suspension as a function of $\phi$. This class
of methods have been reviewed in detail in the first part of our
review.\cite{poon2011} Here we concentrate on a second class of
methods: determining $\sigma_{\rm eff}$ from the inter-particle
potential, $u(r)$. To do so, of course, requires means of measuring
$u(r)$, which is the main topic of this section.

First, however, we briefly review how a knowledge of $u(r)$
can be used to determine an effective hard-sphere diameter,
$\sigma_{\rm eff}$.  Perhaps the simplest approach is to set an effective
hard sphere diameter $\sigma_{kT}$ such that the inter-particle
repulsive energy at this centre-to-centre separation between two
particles is equal to the thermal energy, i.e.
\begin{equation}
\beta u(r=\sigma_{kT})=1, \label{eqkT}
\end{equation} 
where $\beta = 1/k_BT$. A more sophisticated approach, which distinguishes between different potentials with the same $\sigma_{kT}$, is to use the Barker-Henderson effective hard sphere diameter\cite{barker1976}
\begin{equation}
\sigma_{BH}=\intop_0^\infty dr\left[1-\exp\left(-\beta u(r)\right)\right].
\label{eqBH}
\end{equation}
Other, yet more sophisticated mappings, exist, such as that due to Andersen,
Weeks, and Chandler \cite{andersen1971}; this approximation is known
to work well for mapping the static properties of liquids and more recently
structural relaxation time near the glass transition \cite{schmiedeberg2011}.
All of these approaches rely on knowing $u(r)$.  We now review
methods for gaining this knowledge.

\subsection{Direct measurement}
\label{subSectionDirect}

A host of sophisticated  techniques are now available for direct measurement of colloidal forces; integration of the measured force-distance relationship then gives $u(r)$.

In the case of PMMA particles sterically stablised by
poly-12-hydroxyl steric acid (PHSA) `hairs', the interaction may be
inferred via the direct measurement of the interaction potential
between mica surfaces coated by PHSA using the surface-force
apparatus. The results were well described by an inverse power
law, suggesting a reasonably (but \emph{not} absolutely) hard
interaction \cite{bryant2002};  see section \ref{sectionIntrinsic}
for more details.

Other methods for measuring colloidal interactions directly include total
internal reflection microscopy, which measures the force between a
colloid and a glass wall,\cite{bechinger1999}
and atomic force microscopy with a colloid on the tip of the cantilever.\cite{piech2002} The interaction between
two non-index-matched colloids confined to a line can be measured by optical tweezers.\cite{crocker1994}

One attraction of such direct methods is that no \emph{a priori} assumption
need be made about $u(r)$. However, their use requires care. Thus, e.g., in the case of optical tweezers,
relatively small and subtle experimental errors can lead
to the wrong \emph{sign} of the interaction between charged colloids~\cite{baumgartl2005}.

\subsection{Extraction from correlation functions}
\label{subSectionRealSpace}

For a colloid at volume fraction $\phi$, the inter-particle potential, $u(r)$, uniquely determines the system's pair correlation function, $g(r)$, or, equivalently, the structure factor $S(q)$, which is essentially the Fourier transform of $g(r)$. Determining $g(r)$ or $S(q)$ from a given $u(r)$ is, of course, one of the classical problems of liquid state theory.\cite{hansen} In principle, it is also possible to reverse this procedure, and infer $u(r)$ from measured correlation functions. The inversion of $S(q)$ to obtain $u(r)$ has a long history in liquid state physics, and has also been used for colloids.\cite{duits1991,ye1996} A fundamental difficulty with this approach is that, like all inverse problems, this one is ill-conditioned. Essentially, many different forms of interaction can give rise to the same $S(q)$. For our purposes, it is instructive to bear in mind that the $S(q)$ for the inert gases near their respective triple temperatures can be well fitted to the hard-sphere $S(q)$ at an appropriate $\phi$, even though the inter-particular potential under the same conditions is well approximated by a Lennard-Jones form: $u(r) = Ar^{-12} + Br^{-6}$, which includes an attractive part, and a repulsive part that is very far from `hard'. 

\begin{figure}
\centering
\includegraphics[width=5cm,keepaspectratio]{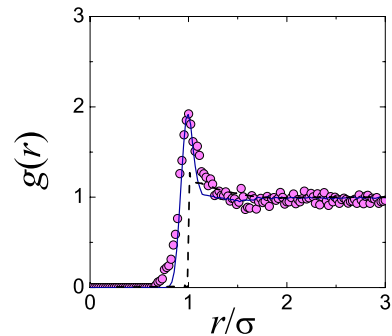} 
\caption{
Attractions in ``hard'' spheres for confocal microscopy. Experimental data are from 
\cite{ohtsuka2008}. The dashed black line is from Percus Yevick theory for hard spheres at $\phi=0.071$ \cite{hansen}.
The solid line is computer simulation data with particle tracking errors and polydispersity added, for a square well
attraction of depth $k_BT$ and range $0.09\sigma$.}
\label{figG} 
\end{figure}

Inverting the real-space pair correlation function, $g(r)$, can also give the interaction potential,\cite{brunner2002,royall2003} and is subject to the same ambiguities, especially when many-body effects are present.\cite{louis2002} But if $g(r)$ can be measured in the limit of vanishing $\phi$, then the inversion to give $u(r)$  is unique:\cite{hansen}
\begin{equation}
\lim_{\phi\rightarrow0}g(r)=\exp\left[-\beta u(r)\right].
\label{eqArd}
\end{equation}
This approach has become possible with the advent of real-space techniques, which allows the determination of $g(r)$ by direct counting. 
\cite{vanblaaderen1995,behrens2001,royall2003} In the limit represented by Eq.~\ref{eqArd}, monodisperse hard spheres give a perfect step function form for $g(r)$, but polydispersity and particle tracking errors would blur the sharpness of the edge at $r = \sigma$, as would a small degree of softness. In Fig. \ref{figG}, we show the $g(r)$ measured in this limit for a putative hard-sphere suspension.\cite{royall2007jcp} It is immediately clear that these particles cannot, in fact, be hard spheres. The peak in the dilute-limit $g(r)$ is due to a short-range inter-particle attraction.\footnote{The step function form of $g(r)$ for dilute hard spheres gives a featureless $S(q) = 1$. Residual attraction or softness therefore shows up much more obviously in real space.} Computer simulations can be used to fit the measured dilute-limit $g(r)$ using a square-well attraction for $u(r)$; importantly, however, a distribution of particle sizes as well as random tracking errors (both modelled by Gaussians) are essential to obtain a good fit. 
Interestingly, it has been argued\cite{louis2001} that
Eq.~(\ref{eqArd}) remains a remarkably accurate approximation at
finite but modest $\phi$.

\subsection{Charged particles}

The application of Eq. \ref{eqArd} makes no {\em a priori} assumption about the form of $u(r)$, and can therefore be applied to any system. For charged systems, if the inter-particle potential is modelled using, e.g. DLVO theory (for which see section \ref{sectionElectrostatics}), then one could access $u(r)$ via measuring the parameters of this theory, viz., an effective charge on the particles, $Z$, and the ionic strength of the solvent.  The definition and therefore determination of $Z$ is highly non-trivial,\cite{russell} and the deduction of ionic strength from conductivity measurements in non-aqueous solvents depends on a number of assumptions\cite{royall2006}. Nevertheless, reasonable results can be obtained \cite{royall2006,royall2003,leunissenThesis}.

\subsection{Accuracy}

All of these methods have limitations. Those based on particle tracking
methods (optical tweezers and extraction from real space) are limited
by the accuracy with which the particle coordinates can be tracked, typically 30-100 nm, or up to $0.05\sigma$ for a $2$ $\mu$m particle. Moreover, there is evidence that tracking errors on two closely approaching colloids
may tend more to the point of contact, rather than being uniformly distributed.\cite{baumgartl2005,royall2007jcp} Furthermore, averaging over many particles in an inevitably polydisperse sample effectively adds a further error to the measurements.\cite{royall2007jcp} 
Comparison with simulations can be used to compensate for some of these errors, though this avenue has been
little explored to date. Finally, at higher concentrations,\cite{royall2003} and in reciprocal space, one needs to make prior assumptions about the form of the inter-particle potential that one wants to determine in the first place.\cite{hansen,duits1991,ye1996} Accurate measurement of $u(r)$ is therefore far from straightforward.

\section{Intrinsic softness}
\label{sectionIntrinsic}

We now turn to review softness in model hard-sphere systems of non-electrostatic origin. There are two generic sources of such `intrinsic' softness: steric stabilisation, and the use of particles with temperature-tuneable size. 

\subsection{Steric stablisation}

\begin{figure}
\centering
\includegraphics[width=9cm,keepaspectratio]{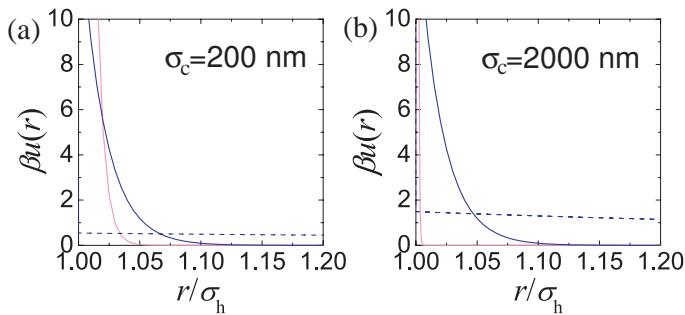} 
\caption{Estimation of effective colloid-colloid interactions in sterically-stabilised PMMA particles: 
(a) $\sigma_H=200$ nm, 
(b) $\sigma_H=2000$ nm.
In both parts, light pink lines denote $u_s(r)$, the interaction due to the steric stabilisation. The dashed blue line in each case represents unscreened weak electrostatic interactions in low dielectric constant solvents (cis-decalin/TCE) calculated for
effective charges $Z=2$ and $Z=16$ for (a) and (b) respectively and a Debye length of $\kappa^{-1}=5000$ nm.
The solid blue line represents typical screened electrostatic interactions: (a) in water (charge  $Z=1700$ and Debye length $\kappa^{-1}=4$ nm), and (b) a density matching mixture of cis decalin and CXB (charge  $Z=500$ and Debye length $\kappa^{-1}=100$ nm). 
} 
\label{figU} 
\end{figure}

A sterically-stabilised particle is shown schematically in Fig. \ref{figRealHard}(a). We have previously\cite{poon2011} reviewed ways to arrive at an effective hard sphere diameter for these particles via mapping to various hard-sphere properties. As already mentioned in section \ref{subSectionDirect}, such mapping is also possible via direct measurement of colloidal forces. This has been done for PHSA-stabilised PMMA particles; the measurements are quite well described by an inverse power law potential with energy scale $\epsilon_s$:\cite{bryant2002}
\begin{equation}
u_{s}(r) \approx \epsilon_s \left(\frac{\sigma_H}{r}\right)^n.
\label{eqUs}
\end{equation}
The relative range of $u_{s}$ depends on the particle size, for
example $n=170$ for particles with a hydrodynamic diameter of $\sigma_H=200$ nm particles, and increases with particle diameter. Likewise the strength of the interaction also depends on the particle size, with $u_{s}(\sigma_H)=146$ $k_{B}T$ for  $\sigma_H=200$ nm. The results of Bryant \emph{et al.} \cite{bryant2002} are replotted in Fig.~\ref{figU}. These quantify what is intuitively obvious, namely, that for a fixed length of stabilising `hairs', larger particles are relatively harder.


\subsection{Microgels}

One main motivation to use soft particles to model hard spheres is that the diameter, and therefore concentration, of these colloids may be temperature tuneable. The most popular systems here are microgel particles. As cross-linked polymer coils, their swelling is related to solvent quality, with the precise degree of swelling controlled by the cross-linking density. There are two varieties: `neutral' microgels, in which a low amount of charges is screened by added salt or is so small as to be negligible, and ionic microgels, which carry (much more) electrostatic charge, for which see section \ref{sectionElectrostatics}. 

Few analytic expressions for microgel-microgel interactions exist. Expressions for polymer-covered flat surfaces have been suggested as a first approximation \cite{berli2000}.
The steric interactions between neutral microgels have been likened to crosslinked polymers, so that a form for the interaction can be obtained provided the density profile of the particle is known.\cite{gottwald2005} Additional interactions in ionic microgels do have analytic forms \cite{denton2003}, which can be added to the steric repulsion, for which an inverse power form is often assumed. 

Irrespective of the absence of well-attested analytical forms for the inter-particle repulsion, swellable colloidal microgels are definitely not hard, and the degree of swelling depends sensitively on the experimental conditions, including possibly the concentration of microgel particles. If the latter is a significant effect, then any $\phi_{\rm eff}$ would become state-dependent. 

We now review a popular system, dispersions of poly
($N$-isopropylacrylamid) (PNiPAM) microgel
particles,\cite{Schild} for which water is a poor solvent at $T
\gtrsim 33^\circ$C, so that at and above this temperature, the
diameter of the particles in an aqueous environment dramatically shrinks. Changing the amount of the cross-linker $N$,$N^\prime$-methylenbisacrylamide tunes how much shrinkage occurs.\cite{senff1999} Sufficiently monodisperse samples crystallise at high volume fractions, giving crystals whose structure has been variously reported as face-centred cubic\cite{Hellweg} or as
consisting of the more or less random stacking of
hexagonally-packed layers.\cite{nguyen2011} Richtering and co-workers have examined the physical properties of PNiPAM microgel suspensions with a variety of probes and discussed their findings in terms of possible mapping onto hard spheres. We mention three aspects.

First, neutron scattering shows that an individual particle has a constant-density core, surrounded by a corona in which the density gradually decreases to zero over a distance that is approximately twice that of the core.\cite{Stieger} From this finding alone, we expect the particles to be significantly soft. 

Next, the structure factors of PNiPAM suspensions at
progressively higher particle concentrations have been measured,
and compared to that of hard spheres.\cite{Stieger2} Using two
different methods of data analysis, it was concluded that the
$S(q)$ of these suspensions could be described adequately within
a hard-sphere framework by assigning $\phi_{\rm eff}$ (or,
equivalently, $\sigma_{\rm eff}$) to the particles at concentrations $\phi_{\rm eff} \lesssim 0.35$. Above this concentration, increasingly large deviations from hard-sphere-like behaviour were observed. Interestingly, in a later paper,\cite{eckert2008} the same group points out that the mapping to hard-spheres at $\phi_{\rm eff}$ was obtained by making one of two assumptions: either that a hard-sphere structure factor (from Percus-Yevick theory) in fact fitted the data, or that the form factor of a single particle\cite{Stieger} determined in the low-concentration limit did not change when the microgel concentration was increased, which is not self-evidently true. 

Finally, Richtering and coworkers investigated the fluid-crystal
coexistence gap.\cite{senff1999,eckert2008} In one study of
PNiPAM particles dispersed in water,\cite{senff1999} a $\phi_{\rm
eff}$ was determined by requiring agreement with the hard-sphere
expression for suspension viscosity at low concentrations,
$\eta/\eta_0 = 1 + 2.5\phi_{\rm eff} + 5.9\phi_{\rm eff}^2$
(where $\eta_0$ is the solvent viscosity). This procedure is problematic because of  the aforementioned
state-dependence of $\sigma_{\rm eff}$ (and therefore of $\phi_{\rm eff}$). Nevertheless, using this mapping, the fluid-solid coexistence gap was found to be $\phi_{\rm eff}^f = 0.59  \leq \phi \leq 0.61 =  \phi_{\rm eff}^m$, i.e. it occurs at significantly higher concentrations than that in hard spheres ($0.494 = \phi_{\rm HS}^f \leq \phi_{\rm HS}^m \leq \phi_m = 0.545$), and is substantially narrower ($\phi_{\rm eff}^m - \phi_{\rm eff}^f  = 0.02$, $\phi_{\rm HS}^m - \phi_{\rm HS}^f = 0.051$).\footnote{The same narrowing of the fluid-crystal coexistence gap has been found in an oil-based system of polystyrene microgels.\cite{iacopini2008}} Comparison with simulations\cite{Kofke} of the freezing of particles interacting via a power-law repulsion $u(r) \propto r^{-n}$ gives $n \approx 13$ for PNiPAM particles with $240 \mbox{nm} \lesssim \sigma_H \lesssim 300$nm. This is considerably softer than the inter-particle potential found\cite{bryant2002} for sterically-stabilised particle of comparable size ($\sigma_c=200$nm), which can be characterised by a power law with exponent $n=170$. 

A different mapping was used in a second study of fluid-crystal coexistence,\cite{eckert2008} in which the the microgel particles are now dispersed in dimethylformamide (DMF, a good solvent chosen for refractive index matching). The freezing concentration of the microgel particles expressed in mass fraction, $\mu = m_{\rm microgel}/(m_{\rm microgel} + m_{\rm solvent})$, was converted to an effective hard-sphere volume fraction by a multiplicative factor, $\phi_{\rm eff} = S\mu$, where the `swelling ratio' $S$ was chosen to yield a freezing volume fraction of $\phi_{\rm eff}^f = 0.494$. Interestingly, this procedure gave $\phi_{\rm eff}^m \approx 0.55$ for the point at which 100\% crystallisation should occur, consistent with hard-sphere behaviour. This agrees with the findings by another group of a recent imaging study using larger PNiPAM particles dispersed in aqueous medium, \cite{nguyen2011} but contrasting strikingly with the previous finding by the same group of a significantly narrower coexistence gap.\cite{senff1999} The same study found that the collective diffusion of these microgel particles dispersed in DMF and their hydrodynamic interactions could {\em not} be well described by any mapping to hard spheres. 

These studies illustrate some of the difficulties associated with mapping microgels to hard spheres. An additional issue is that of polydispersity. Both softness in the inter-particle potential\cite{Kofke} and polydispersity\cite{wilding2011} affect the width of the coexistence gap, so that the effect of these two quite distinct physical factors may be difficult to disentangle. Fortunately, microgels can be synthesised with polydispersities as low as $\sim 1\%$ before swelling, so that perhaps the polydispersity effect can be neglected in the first approximation (cf. the very small effect 1\% polydispersity on the miscibility gap of hard spheres\cite{wilding2011}). 

A variant of the `canonical' microgel has been synthesised and
characterised by Ballauff and
co-workers\cite{Ballauff2006,Ballauff2008} consisting of a hard
polystyrene core onto which is grafted a network of cross-linked
PNiPAM, so that the swollen shell has approximately the same
dimensions as the core radius ($\approx 50$~nm). To determine
$\phi_{\rm eff}$, a core volume fraction $\phi_c$ was first
measured by conversion from the particle mass fraction using the
density of polystyrene. The hydrodynamic diameter of the
particles, $\sigma_H$ was then determined using dynamic light
scattering; this was later confirmed to be very close to the
diameter of the outer corona visible in cryo-transmission electron
microscopy (cryo-TEM) images,\cite{Ballauff2008} which also gave
the core diameter $\sigma_c$, and the polydispersity of the
core-shell diameter distribution ($\approx 9\%$). Finally, using $\phi_{\rm eff} = \phi_c(\sigma_H/\sigma_c)^3$, the fluid-crystal coexistence gap at 21$^\circ$C was  found at $0.483 \pm 0.007 < \phi_{\rm eff} < 0.546 \pm 0.007$, which is, within experimental uncertainties, very close to the hard-sphere interval of 0.494 to 0.545.

It is interesting to analyse these findings further. As the authors themselves\cite{Ballauff2008} have pointed out, if $\phi_{\rm eff}^f$ is rescaled to exactly 0.494, then melting occurs at 0.556.  This gives a coexistence gap wider than in perfect hard spheres. If this does not reflect experimental errors, then the situation is somewhat unusual -- the most common `culprits', polydispersity and softness in the repulsive potential, both {\em narrow} the coexistence gap [Fig. \ref{figMiscibility} (a) and (b)]. 
However, short-range attraction in the potential has the opposite effect of widening the coexistence gap if the polydispersity\footnote{Experimentally, for hard spheres above a critical polydispersity, short-range attraction appears not to widen the coexistence gap.\cite{Siobhan}} is low enough\cite{gast1983,lekkerkerker1992,eliott1999}. 
On the other hand, 9\% polydisperse hard spheres should be at or beyond the experimental limit of crystallisation,\cite{pusey1987} probably due to the onset of multiple solid phase coexistence in the phase diagram,\cite{Zaccarelli} which requires long-range particle motion for fractionation. That crystallisation was still observed in these PNiPAM samples to give a coexistence gap wider than that of hard-spheres underlines the the lack of complete understanding of microgel physics. 

More recently, Ballauff and co-workers studied in detail the rheology of a 17\% polydisperse suspension of their core-shell particles at and near the glass transition (found to occur at $\phi_{\rm eff} = 0.640$), and compared their data with mode-coupling theory (MCT) calculations for hard spheres.\cite{Ballauff2009} At this polydispersity crystallisation was inhibited, but this introduced an extra level of complexity into calculating $\phi_{\rm eff}$. The authors relied on the fact that the high-frequency viscosity had been found to be relatively insensitive to polydispersity,\cite{cheng} and mapped the high polydisperse system onto a less polydisperse system in which $\phi_{\rm eff}$ had already been calibrated according to the procedure explained above. A large measurement of agreement with MCT predictions has been found. We note in this connection that the comparison with MCT mostly relies on a relative measure of the distance to the glass transition $\phi_g$: $\epsilon = (\phi - \phi_g)/\phi_g$, so that the work is perhaps less vulnerable to systematic or statistical uncertainties in arriving at $\phi_{\rm eff}$. 

\subsection{Star Polymers}

In the spectrum of hard to soft colloids, microgels
occupy a middle niche. Towards the soft end of this
spectrum, we encounter star polymers; when the number of `arms'
(or `functionality, $f$) is high enough, these can
be considered soft colloids, with decreasing softness  as $f$
increases.  Thus, for example, stars with $f > 34$ are predicted
to crystallise \cite{Lowen2000}. But the structures of the crystal
phases differ from the crystals nucleated from hard spheres (random
stacking of hexagonal layers). Moreover, the effective potential
between two particles at close approach is logarithmic,\cite{Witten}
which is significantly softer than the power-law forms encountered
in this review so far. Thus, although in a mixture of (large)
hard colloids and (smaller) star polymers, 32-arm stars were
found to behave almost hard-sphere like vis-\`a-vis the hard
colloids,\cite{Poon2001} it is unlikely that pure star polymer
suspensions can be used as models of hard spheres.

\section{Electrostatics}
\label{sectionElectrostatics}

It is now generally accepted that immersion of a colloid in a liquid medium {\em always} gives rise to some degree of charging of the particles. This is a potential source of softness that should always be considered in experiments. If charged groups are present on the surface of the particles, entropy inevitably favours at least a certain degree of dissociation; even when charged surface groups are absent, adsorption of charged species from the medium will give rise to charged particles.   

The treatment of electrostatic interactions on the mean-field, or linearised Poisson-Boltzmann (PB), level is largely adequate for our purposes. This is especially true for non-aqueous systems,\cite{royall2005,royall2006} in which the energetic penalty of ionisation
is high, so that ion densities are low, and multivalent ions can be safely neglected.

The linearised PB theory is incorporated into the Derjaguin-Landau-Verwey-Overbeek (DLVO) theory\cite{verwey1948} to describe the interaction between charged colloids. The original DLVO potential
consists of van der Waals (vdW)
and electrostatic components. We are
primarily interested in situations in which sterically-stabilised
particles become charged, so we will assume that the steric
repulsion is adequate to render the vdW component negligible. Instead, we will consider
an inter-particle potential consisting of a steric repulsion,
$u_s(r)$, and an electrostatic interaction, which in linearised PB theory has an Yukawa form, $u_Y(r)$:
\begin{eqnarray}
u(r)&=&u_{s}(r)+ u_{Y}(r),
\label{eqU}\\
u_{Y}(r)& =& \epsilon_{Y}\frac{\exp [-\kappa ( r-\sigma_c ) ]  }{r/\sigma_c}.\label{eqYuk}
\end{eqnarray}
Here, the contact potential given by
\begin{equation}
\beta\epsilon_{Y}=\frac{Z^{2}}{(1+\kappa\sigma_c/2)^{2}}\frac{\lambda_{B}}{\sigma_c}
\label{eqEpsilonYuk}
\end{equation}
where, $Z$ is the colloid charge, and the inverse Debye screening length is given by $\kappa$=$\sqrt{4\pi\lambda_{B}\rho_{ion}}$, where
$\rho_{ion}$ is the number density of monovalent ions. The Bjerrum length 
\begin{equation}
\lambda_{B}=\beta e^{2}/(4\pi\epsilon_{0}\epsilon_{r}), \label{eqBjerrum}
\end{equation}
is the distance at which the interaction energy between two
electronic charges is $k_B T$, where $e$ is the electronic
charge, $\epsilon_{0}$ the permittivity of free space, and
$\epsilon_{r}$ the dielectric constant.  While this form of the
electrostatic interaction is only valid in the range that linearised
PB theory holds (weak electrostatic interactions),
higher charging can also be treated with a Yukawa interaction
by using a  \emph{renormalised} charge that is smaller than the
physical charge on the particles.\cite{alexander1984}

\subsection{Strong charging in aqueous solvents}

Water has an unusually large zero-frequency dielectric constant, $\epsilon_r \approx 80$ at around room temperature, giving a Bjerrum length of 0.7nm. Since $\lambda_B$ is comparable to the size of small ions, many ionic salts are readily soluble in water. This in turn means
that the Debye length is easily controlled down to a few nanometres. 
For $\sigma = 400$nm and $Z=1700$ in water with a $\kappa^{-1}
=  4.0$nm, Eqs. \ref{eqYuk}, \ref{eqEpsilonYuk} and
Eqs. \ref{eqkT} and \ref{eqBH} give $\sigma_{kT}=1.052\sigma_c$
and $\sigma_{BH}=1.062\sigma_c$. Thus, screened charged colloids
in water very often approximate hard spheres reasonably if they
are either sterically stabilised or the short $\kappa^{-1}$ is
still large enough to render vdW attractions irrelevant. Nevertheless,
some aspects of the behaviour of such colloids may still be non-hard-sphere-like. Thus, Piazza~\emph{et
al.}\cite{piazza1993} found that, although the equation of state
(EOS) of their screened charged colloids approximated
the hard sphere EOS at moderate to high $\phi$, at low $\phi$,
the barometric distribution in the earth's gravitational field,
$\phi(z) = \phi(0)e^{-z/z_0}$, was {\em not} observed, with
the measured height distribution showing a more slowly decaying
`tail'. This is related to a decoupling of ions and colloids,
leading to a macroscopic electric field \cite{vanRoij2003}.

\subsection{Weakly-charged small particles in non-aqueous media}
\label{sectionNonAqueous}

\begin{figure}
\centering
\includegraphics[width=8cm,keepaspectratio]{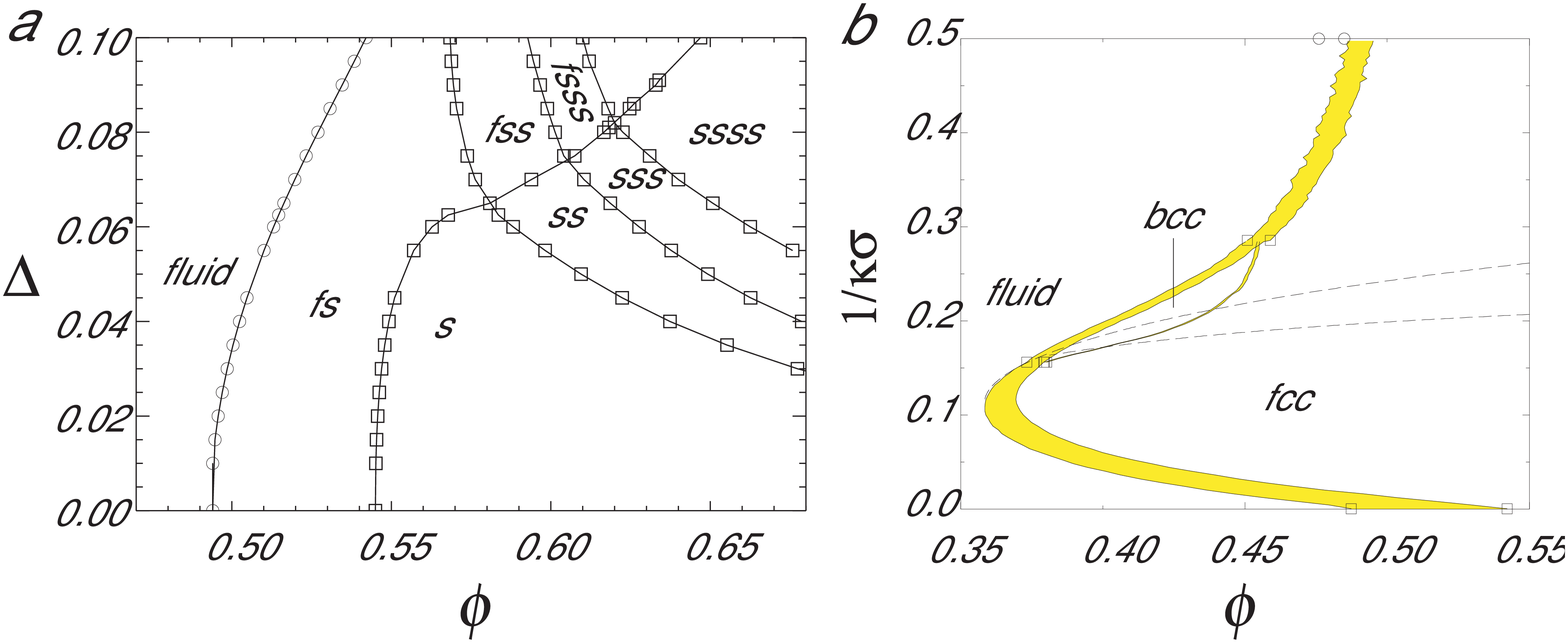} 
\caption{(a) The theoretical phase diagram of hard spheres at different polydispersities, $\sigma$. F = fluid, S = (crystalline) solid; thus FSS denotes fluid-solid-solid coexistence. Replotted from Wilding and Sollich \cite{wilding2011}. (b) Phase diagram of hard-core Yukawa particles, from \cite{hynninen2003} in the absolute volume fraction - Debye length $(1/\kappa\sigma)$ plane. Here the contact potential $\epsilon_{Y}=8k_{B}T$. In the case of zero Debye length, the hard sphere limit is recovered. Replotted from Hynninen and Dijkstra \cite{hynninen2003}.}
\label{figMiscibility} 
\end{figure}

In non-aqueous solvents, charging is profoundly altered. The reduced
dielectric constant, $\epsilon \lesssim 10$, leads to $\lambda_{B}\sim10-40$
nm. With ionic sizes in the range of 1 nm, one expects strong coupling between oppositely-charged ions and and therefore little dissociation of surface groups. It was therefore a long-held assumption that electrostatics could be safely neglected. However,
the work of the van Blaaderen group~\cite{yetiraj2003} and others have shown that some electrostatic charging \emph{always} occurs. 

It seems that the degree of charging in many systems\cite{trizac2002, roberts2007, royall2003, leunissenThesis, alexander1984} can be described by the rule of thumb $Z \lambda_B / \sigma \approx 6$.
Thus, the system used by Pusey and van Megen,\cite{pusey1986} sterically-stabilised PMMA ($\sigma \approx 700$nm) in a mixture of cis-decalin and carbon disulphide ($\epsilon_r = 2.64$, $\lambda_B \approx 20$nm), can be expected to be charged to some extent, as later work on a similar suspension seems to confirm.\cite{roberts2007} 

Nevertheless, when the particles are small enough, such charging can often be ignored. Figure \ref{figU}(a) (dashed red line) shows $u_Y(r)$ for a charge of $Z=2$ on the surface of $\sigma_H=200$~nm particles (corresponding to $Z \lambda_B / \sigma \approx 6$) in a dispersion medium with a Debye length of $\kappa^{-1} = 5\mu$m. The measured steric repulsion for sterically-stabilised PMMA particles of this size\cite{bryant2002} is also shown. It is clear that for all relevant length scales in this situation, $u_Y(r) \ll k_B T$. These particles can plausibly be considered hard spheres. 


\subsection{Particles for confocal imaging}
\label{subSectionConfocal}

The most popular hard-sphere model system to date for confocal microscopy is sterically-stabilised PMMA, because of the possibility of using a solvent mixture for simultaneous refractive index and density matching. For accurate determination of coordinates, particles with $\sigma_H \gtrsim 1\mu$m are required. The density-matching solvents used are typically halogenated hydrocarbons. The use of these solvents has a number of undesirable side effects. They tend to swell the particles much more aggressively than non-halogenated hydrocarbons, and they can damage the fluorescent dye molecules included in the particles for laser confocal microscopy. They also lead to significant leves of electrostatic charging. 

Four commonly used halogenated solvents are 
cycloheptyl bromide (CHB), cyclohexyl bromide (CXB), tetrachloroethylene
(TCE), and carbon tetrachloride.\footnote{Note that carbon tetrachloride is a suspected
carcinogen.} In these low dielectric constant solvents (e.g., $\epsilon_r = 7.9$ for CXB), colloids suitable for
confocal microscopy ($\sigma_{H}\sim2$ $\mu$m) acquire a charge
of $Z\sim 100-500$. The Debye length in a density-matching CXB-cis
decalin mixture can run to microns as the ionic
strength (due predominantly to solvent self-dissociation) can be as low as
$10^{-10}$ M, \cite{yetiraj2003} which is much lower than the ionic concentration in pure water ($10^{-7}$
M). 

The charge on the colloid, although much lower than what
can be expected on similar sized particles in water, is
now almost unscreened, which can lead to very long-ranged
and strong interactions ($\epsilon_{Y}\gtrsim$100 $k_{B}T$).
These interactions can and do vary from sample to sample, as the
ionic strength in CXB and CHB varies from batch to batch, and as a
function of time.\cite{royall2006,royall2003,leunissenThesis}
Colloidal crystallization has been found in some
cases\cite{gasser2001} at $\phi\approx0.4$ (as compared to
hard spheres at $\phi^f_{\rm HS} = 0.494$), but at least one experiment
saw crystallization at volume fractions as low as
$\phi\sim0.01$;\cite{yetiraj2003}  often these
crystals were body-centered-cubic (bcc) in contrast to
hard-sphere crystals which are random-hexagonal-close-packed (rhcp) or fcc.
Furthermore, for some batches of CXB, the colloid charge can change
with particle concentration, leading to strongly $\phi$-dependent
interactions, and even to re-entrant melting.\cite{royall2006}
The use of PMMA particles in halogenated solvents for confocal
microscopy is therefore problematic.

The situation can be improved somewhat by the use of salts to screen the charges. The problem with this approach is that salts soluble in these solvent mixtures such as tetrabutyl ammonium bromide (TBAB)\cite{yetiraj2003}
are soluble only to around 260 nM.\cite{leunissenThesis,royall2006}
This results in a Debye length of $\kappa^{-1} \approx 100$~nm. Although this is
substantially less the diameter of imageable colloids ($\sigma_H \gtrsim 1\mu$m), it is not negligible and
 a noticeable degree of softness will likely
result, Fig. \ref{figU}(b). On the other hand, since the majority of ions now come from
the salt, the ionic strength and therefore the colloid-colloid effective
interactions will likely be reasonably independent of $\phi$.

Another possibility is to use lower dielectric
constant solvents such as TCE and CCl$_{4}$ ($\epsilon_r=2.5$ and $2.24$ 
respectively). Lower $\epsilon_r$ increases $\lambda_B$, Eq.~\ref{eqBjerrum}. The rule of thumb for estimating the degree of charging, viz., $Z \lambda_B / \sigma_c \approx 6$, therefore predicts a lower $Z$. However, both of these solvents are strongly
absorbed by PMMA, and can lead to a volume swelling of $\gtrsim 40$\%.\cite{ohtsuka2008} Unless the swelling is very closely monitored and characterised, it becomes a serious source of potentially large systematic errors,\cite{poon2011} because $\phi \propto \sigma^3$. Solvent absorption also increases the density and refractive index of the particles. Thus, one of the initial attractions of using such halogenated solvents is lost - without swelling, adding one of these solvents can density match nearly exactly but also (fortuitously) nearly match the refractive index. With significant absorption, more TCE (say) than is needed for index matching has to be added to achieve density matching. Unless a third solvent is used to re-achieve index matching (which itself may need to further swelling), a turbid sample results. 

Such turbidity not only degrades image quality, but can also give rise to significant vdW attraction. For example,  Fig. \ref{figG}, which shows the measured $g(r)$ of a $\phi = 0.071$ suspension of sterically-stabilised PMMA particles in a density-matching mixture of cis-decalin and TCE.\cite{ohtsuka2008} The pronounced peak at touching immediately alerts us to the presence of inter-particle attraction, as simulations confirm.

\section{A cautionary worked example}

\label{sectionCaseStudy}

Since a well-attested analytical form for the interaction between charged colloids is available, Eqs. (\ref{eqU})-(\ref{eqBjerrum}),  we close by presenting an `exact' comparison between the simulated phase behaviour of such particles, Fig. \ref{figMiscibility}, and various ways of mapping their behaviour to hard spheres. This comparison illustrates how careful one must be in drawing conclusions from such mapping, even in the limit when inevitable experimental uncertainties (e.g. due to polydispersity or slightly soft steric-stabilising `hairs') are negligible. 

Consider charged hard spheres, so that $u_s(r)$ in Eq.~(\ref{eqU})
is the perfect hard-sphere repulsion. We model scenarios that may
reasonably represent sterically-stabilised PMMA used in confocal
imaging, and take $\sigma = 2$~$\mu$m, and a Debye screening
length of $\kappa^{-1} = 100$~nm (so that $\kappa\sigma = 20$).
The latter is a round figure chosen to correspond roughly to CXB
with the maximum possible amount of
dissolved TBAB (260~nM). Consider two colloid charges $Z = 500$ and $Z = 100$. The former is consistent with the rule of thumb, $Z\lambda_B/\sigma \approx 6$, while the latter is 5 times lower than predicted by this empirical relation. These charges have been reported in different studies\cite{royall2003,royall2006,sedgwick2004,campbell2005} for nominally identical PMMA particles used for confocal imaging dispersed in a density-matching mixture of cis-decalin and CXB. Equation (\ref{eqYuk}) gives $\epsilon_{Y} \approx 10 k_BT$ and $\approx 0.5 k_BT$ for these two charges respectively.

\begin{figure}
\centering
\includegraphics[width=8cm,keepaspectratio]{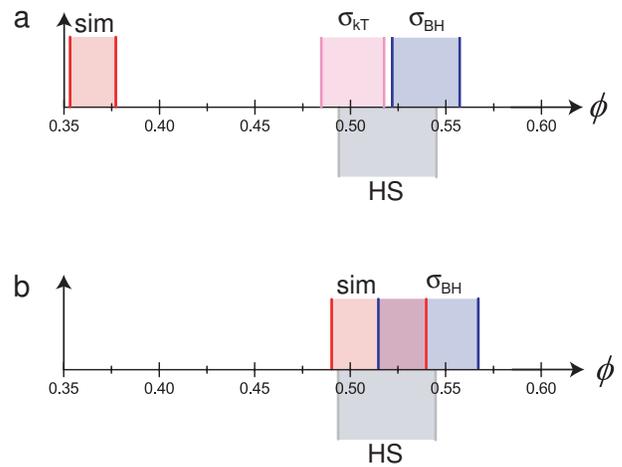} 
\caption{Mapping the phase behaviour of two hypothetical monodisperse charged hard sphere colloids (with parameters based on real systems -- see text for details) to pure hard spheres. The hypothetical particles and solvent have the following properties: $\sigma = 2\mu$m, $\kappa^{-1} = 100$nm. The particles have two different charges (a) $Z = 500$, (b) $Z = 100$. In each case, differently shaded and delimited regions denote the fluid-solid coexistence gap of pure hard spheres (grey, `HS'), and from: simulations\cite{hynninen2003} (red, `sim'), mapping using Eq.~(\ref{eqU}) (lilac, `$\sigma_{kT}$'), and mapping using Eq.~(\ref{eqBH}) (blue, `$\sigma_{BH}$'). In (b), the result of mapping using Eq.~(\ref{eqU}) does not change the coexistence region from that predicted by `sim', and is not shown separately.}
\label{figMapping} 
\end{figure}

Taking these parameters, the simulation results\cite{hynninen2003}  replotted in Fig. \ref{figMiscibility}(b) can be used to determine freezing and melting for our hypothetical systems.  These are delimited by red lines in Fig. \ref{figMapping} for (a) $Z=500$ and (b) $Z=100$. Note first that these values of not-very-large surface charge, variously reported in the literature for nominally very similar PMMA colloids, in fact pertain to rather large differences in the freezing/melting transitions, both in terms of absolute values of the transitions points and in terms of width of the coexistence region. Thus, some kind of `mapping' is clearly necessary if we are to use either system to model hard spheres meaningfully.


We proceed to calculate $\sigma_{\rm eff}$ for the particles
represented by these two parameter sets using either
Eq.~(\ref{eqkT}) or Eq.~(\ref{eqBH}), and Eqs.
(\ref{eqU})-(\ref{eqBjerrum}). If mapping is to be helpful, we should expect that once we have transformed $\phi$ to $\phi_{\rm eff}$, freezing and melting of the two system should occur at approximately the hard-sphere values, viz., 0.494 and 0.545. 

Fig. \ref{figMapping}(a) shows that for the case $Z=500$,
Eq.~(\ref{eqkT}) predicts freezing at $\phi_{kT}^f=0.484$, just
0.010 from the `correct' value.\footnote{To put this difference
in context, note that it is smaller than other sources of errors
inherent in measuring $\phi$.\cite{poon2011}.} However, the width
of the coexistence gap is reduced, which is the result of the
softness of the screened electrostatic (Yukawa) repulsion. Since
the mapping involves scaling $\phi$ by  constant, $\phi_{\rm eff}
= (\sigma_{\rm eff}/\sigma)^3 \phi$, it preserves the {\em
relative} coexistence gap, so that the melting concentration,
$\phi_{kT}^f$ is significantly underestimated. Conversely, the
Barker-Henderson treatment gives a relatively accurate estimate of $\phi_{kT}^m$, but rather significantly overestimates $\phi_{kT}^f$. In the case of $Z=100$,  since $\epsilon_Y < k_B T$, $\sigma_{kT} = \sigma$, so that Eq.~(\ref{eqkT}) maps perfectly onto hard spheres. In this case, the Barker-Henderson approach does the worse job, even when compared to the raw (unmapped) coexistence gap given by simulations of the bare Yukawa interaction. We note that since the coexistence gap varies with the interaction, mapping to the true hard sphere volume fraction at freezing inevitably gives an erroneous melting volume fraction unless the colloids are absolutely hard.

Figure \ref{figMapping} therefore shows that the different methods of mapping to hard spheres
do \emph{not} give the same result. In practice, of course, the inter-particle potential is {\em not} exactly known, and polydispersity is inevitable. 
What is clear from the worked example summarised in Fig. \ref{figMapping} is that even in the `ideal' case of monodisperse spheres with an exactly-known inter-particle interaction, mapping to hard spheres is system- and approach-specific. In practice, of course, polydispersity introduces significant uncertainties, and the Debye length is often not determinable to high accuracy. Moreover, we stress that the values of $Z=500$ and $Z=100$ are taken from experiments on \emph{nominally identical} systems. Thus, conclusions derived from any `mapping to hard spheres', e.g. comparison of nucleation rates at nominally equivalent state points in the coexistence gap, must be treated with significant caution.  

\section{Conclusions}
\label{sectionConclusions}

We set out on a quest for colloids that mimic as closely as possible the ideal hard sphere. It seems that, to date, small (say $\sigma \lesssim 200$nm) sterically PMMA particles dispersed in index-matched hydrocarbons come about as close as possible to this ideal. Larger PMMA particles suitable for confocal microscopy require density matching to minimise sedimentation; but the solvents used inevitably induce a degree of charging that is difficult to screen out entirely using salts. The resulting soft, screened Coulomb inter-particle repulsion can be satisfactorily modelled on the mean-field level, however the interaction parameters vary hugely for nominally identical systems. The use of microgels such as PNiPAM, either on their own or as `shells' on hard `cores', has become popular, because their volume fraction can conveniently be tuned by temperature. Their inevitably soft mutual interaction has proven harder to model in a generic, analytical form. In both cases, the softness necessitates the use of a $\phi_{\rm eff}$ to map onto hard spheres. We have reviewed various ways of performing this mapping, with or without the benefit of knowledge of the inter-particle potential $u(r)$. None seem entirely satisfactory. 

We end by making two further observations. First, we widen the scope of our enquiry from charged particles and microgels to other kinds of non-hard inter-particle interaction, and ask what requirements should be satisfied before one may fruitfully embark on the exercise of `mapping' to hard spheres. We suggest the the minimum conditions to be satisfied are:\\
(1) the absence of any attractive interaction, so that the equilibrium physics is dominated by entropic effects; \\
(2) crystals in a sufficiently monodisperse dispersion at high concentrations consist of the stacking of hexagonal layers.\\
The first criterion highlights the importance of refractive
index matching to minimise the ubiquitous vdW interaction. The
second criterion explains why charged hard particles with
$\kappa \sigma \gtrsim 6$, Fig.~\ref{figMiscibility}, and
microgels\cite{Hellweg,nguyen2011} are suitable candidates
for mapping to hard spheres, but star  polymers are
not.\cite{Lowen2000} 

Secondly and finally, we point out that new developments in particle synthesis may yet produce $\mu$m-sized colloids that can be index {\em and} density matched using solvents or solvent mixtures that do not bring about charging and minimal swelling. 
Such development will be most welcome for the community of scientists wishing to use colloids to test fundamental theories of many-body physics via the hard-sphere model system. 

Of course, soft particles, from microgels to star
polymers and beyond, are fascinating systems in their own
right.\cite{vlassopoulos2012}  Furthermore, it is reasonable to
enquire how much deviation from perfect hard spheres is acceptable,
and the answer of course depends on what one wishes to study.
Section \ref{sectionCaseStudy} shows that the location of colloidal
phase boundaries can vary in nontrivial and qualitative ways from
hard spheres. However, it is plausible that slight softening
may only have slight changes in, for example, the structural relaxation time
\cite{schmiedeberg2011}.


Our point,
then, is that while particle interaction details may not matter
in some cases (such as the pair structure of dense liquids), 
there are plenty of cases where the behaviour of the system depends strongly on 
both the volume fraction $\phi$ \emph{and} the interparticle interactions, therefore accurate knowledge of the both is essential. Crucially, the interactions may well be known to even less precision that the absolute volume fraction $\phi$, which we have already
argued is knowable only to 3-6\%.

\section*{Acknowledgments}

We thank P. Bartlett, J. C. Crocker, R. Evans, D. Frenkel,
M. Fuchs, A. A. Louis, P. N. Pusey, H. Tanaka and A. van
Blaaderen  for helpful discussions over many years,
and P. Sollich and M. Dijkstra for the data in Fig.~\ref{figMiscibility}. CPR is
funded by the Royal Society. WCKP held an EPSRC Senior Fellowship
(EP/D071070/1), and thanks Universit\"at Konstanz for hospitality,
during which part of this work was done. ERW was supported by NSF
grant NSF CHE-0910707.


\end{document}